\documentclass[aps,showpacs]{revtex4}
\usepackage{amssymb}
\usepackage[dvips]{graphicx}
\usepackage[english]{babel}
\usepackage{indentfirst}
\usepackage{amsxtra}
\usepackage{amsmath}
\usepackage{supertabular}
\usepackage{multirow}
\usepackage [mathcal]{eucal}
    \usepackage{etex}          
    \reserveinserts{18}        
    \usepackage{morefloats}    

\def\beq{\begin{equation}}
\def\eeq{\end{equation}}
\def\beqn{ \begin{eqnarray} }
\def\eeqn{ \end{eqnarray} }
\def\s1s2{{ \boldsymbol{\sigma}(1) \cdot \boldsymbol{\sigma}(2) }}
\def\t1t2{{ \boldsymbol{\tau}(1) \cdot \boldsymbol{\tau}(2)  }}

\newcommand{\doc}{{\cal D}}

\newcommand{\coh}{{\cal C}}
\newcommand{\nst}{$N^*$~}

\newcommand{\epy}{$\omega_{\rm sep}$} 

\newcommand{\bsigma}{\mbox{\boldmath $\sigma$}}
\newcommand{\btau}{\mbox{\boldmath $\tau$}}
\newcommand{\half}{\frac{1}{2}}

\newcommand{\br}{{\bf r}}

\newcommand{\threej}[6]{ \left( \begin{array}{ccc}
                               #1 & #2 & #3 \\
                               #4 & #5 & #6 
                             \end{array}
                        \right) }

\usepackage[usenames]{color}
\usepackage{ulem}

\begin{document}
\noindent
\title{Pygmy and giant electric dipole responses of medium-heavy 
nuclei in a self-consistent Random Phase Approximation approach with
finite-range interaction}
\author{G. Co', V. De Donno}
\affiliation{Dipartimento di Matematica e Fisica ``E. De Giorgi'',
 Universit\`a del Salento and,
 INFN Sezione di Lecce, Via Arnesano, I-73100 Lecce, ITALY}
\author{M. Anguiano and A. M. Lallena}
\affiliation{Departamento de F\'\i sica At\'omica, Molecular y
  Nuclear, Universidad de Granada, E-18071 Granada, SPAIN}
\date{\today}

\bigskip

\begin{abstract}
The pygmy dipole resonance (PDR) is studied in various medium-heavy
nuclei by using a Gogny interaction in a self-consistent Hartree-Fock
plus Random Phase Approximation method. We compare the details of the
PDR structure with those of the giant dipole resonance (GDR). In the
PDR protons and neutrons vibrate in phase, and the main contributions
are given by particle-hole excitations involving the neutrons in
excess. On the contrary, in the GDR protons and neutrons vibrate out
of phase, and all the nucleons are involved in the excitation. The
values of the parameters we used to define the collectivity of an
excitation indicate that the PDR is less collective than the GDR.  We
also investigate the role of the residual interaction in the
appearance of the PDR and we find a subtle interplay with the shell
structure of the nucleus.
\end{abstract}

\bigskip
\bigskip
\bigskip


\maketitle

\section{Introduction}

The pygmy dipole resonance (PDR) is an electric dipole excitation
located at energies close to the nucleon emission threshold, and
called pygmy since its strength is much smaller than that of the well
studied giant dipole resonance (GDR) \cite{paa07,kre09}. The reasons
of the interest in the presence of the PDR can be grouped in three
categories.

A first type of motivations is related to nuclear astrophysics. The
presence of a PDR in neutron rich nuclei, would increase the neutron
capture cross section by orders of magnitude \cite{gor98,dao12}, and,
consequently, also the rate of nucleosynthesis r-process.  This effect
could clarify the present discrepancy between simulations and observed
abundance of medium-heavy nuclei in the solar system \cite{arn07}.

A second type of reasons inducing the interest in the PDR is due to
the relation between the presence of this resonance and the nuclear
matter symmetry energy, and consequently, the connection  
with the neutron skin, the static dipole polarizability 
and the values of the Landau's parameters of the effective
nucleon-nucleon interaction \cite{boh81,kri82,car10,rei10,ina11}.

Finally, the third set of reasons to be interested in the PDR is
related to a genuine curiosity about this new type of nuclear
excitation. There are many aspects of this excitation to be
clarified. For example, if the PDRs are present in all the medium and
heavy nuclei or if this type of excitation is peculiar of neutron rich
nuclei only. Another question to be answered is whether these
resonances are the low energy tail of the GDR or if they represent a
different type of excitation. It also remains to be clarified if the
PDRs are produced by collective nuclear motions or if they are
generated by single particle (s.p.) excitations related to the
specific shell structure of nuclei with neutron excess, i.e. $N>Z$.  

Our work has been mainly inspired by this last type of motivations.
We have investigated the presence and the structure of the PDR in
various medium-heavy nuclei, representative of different regions of
the nuclear chart. The theoretical tool we have used in our study is
the Random Phase Approximation (RPA) theory.

The first results of our study have been presented in
Ref. \cite{co09b} where we used a phenomenological RPA approach.  The
s.p. basis was generated by using a Woods-Saxon well. The parameters
of this well were fixed, for each isotope chain, in order to reproduce
some empirical properties of the most investigated doubly magic
nucleus of the chain. This implied that s.p. wave functions and
energies were the same for each isotope of the chain.  The effective
nucleon-nucleon interaction used in those RPA calculations was a
density dependent zero-range Landau-Migdal force, whose parameters
were chosen to reproduce the energies of the low-lying $3^-$ states
and the centroid energies of the GDR in $^{16}$O, $^{40}$Ca and
$^{208}$Pb nuclei.  The main result of that investigation was the
observation of dipole strength at energies lower than those of the GDR
that increased with the neutron excess. In that region proton and
neutron dipole transition densities were in phase, while they were out
of phase in the GDR region. 

From the theoretical point of view the main limitations of the
previous investigation were the large number of parameters to be
selected, and the rigid use of a s.p. basis within a given set of
isotopes.  We overcome these difficulties in the present study by
adopting a self-consistent Hartree-Fock (HF) plus RPA approach.
The effective interaction used in HF calculations to generate 
the s.p bases is also used in the RPA calculations.

Recently, we have proposed two parametrizations of the finite-range
Gogny interaction containing also tensor terms \cite{ang11,ang12}. 
However, since the presence of these terms is relevant only in magnetic
excitations \cite{co12b}, 
in the present study we used 
the well tested D1S \cite{dec80} and the more modern D1M \cite{gor09}
Gogny forces that do not include tensor terms.  
The parameters of these two interactions have
been selected to reproduce some global nuclear properties, therefore
each interaction is built to be used to investigate every nucleus in
each region of the nuclear chart. Contrary to what has been done in
Ref. \cite{co09b}, in the present approach each nucleus has its own
s.p. basis.  The two parametrization produce similar results whose
differences are not large enough to induce to different
conclusions. For this reason we have restricted our presentation to
the results obtained with the D1M parametrization only.

For our investigation we have selected a variety of nuclei with closed
sub-shells and, therefore, with spherical symmetry. In this way, not
only, we have avoided deformation problems, but we have also minimized
the eventual pairing effects. We have chosen isotope chains containing
at least a nucleus where experimental evidence of presence of PDR has
been found: $^{22}$O \cite{lei01}, $^{48}$Ca \cite{har04}, $^{68}$Ni
\cite{wie09}, $^{132}$Sn \cite{adr05} and $^{208}$Pb
\cite{rye02,pol12}.

The paper is organized as follows. In Sec. \ref{sec:model} we briefly
present the main features of our self-consistent HF+RPA model. In Sec. 
\ref{sec:resu} we present, and discuss, our results and in 
Sec. \ref{sec:conc} we draw our conclusions. 

\section{The model}
\label{sec:model}
The first step of our procedure consists in constructing the
s.p. basis by solving the HF equations. The technical details
concerning the solution of the HF equations can be found in
Refs. \cite{co98b,bau99}. In the second step, the wave functions
obtained in the HF calculation are used to solve the RPA equations by
considering, without approximations, the fact that the main part of
the s.p. wave functions above the Fermi surface lies in the
continuum. We have labeled continuum RPA (CRPA) the results obtained
in these calculations. The technique used to solve the CRPA equations,
which takes care of both direct and exchange matrix elements in the
RPA, is based on expansions on a basis of Sturm-Bessel functions. The
details of this approach can be found in
Refs. \cite{don08t,don11a}. The number of the integration points and
of the Sturm-Bessel functions to be used is easily determined by
observing the numerical stability of the results (see
\cite{don08t,don11a}). Apart from that, CRPA calculations are
parameter free because the interaction is fixed.

Unfortunately, CRPA calculations have two drawbacks. A first one is
that the CRPA equations are formulated with very involute expressions
which make very difficult to disentangle the role of the various
ingredients of the calculations, for example the relevance of specific
particle-hole (p-h) excitations. The second problem is that the
calculations are numerically very involved and, for large nuclei our
technique suffers of numerical instability. For these reasons,
together with the CRPA calculations, we carried out RPA calculations
where only a discrete set of s.p. wave functions is used. We call
discrete RPA (DRPA) this last type of
calculations.  The discrete configuration space is obtained in the HF
calculation when the iterative procedure has reached the minimum value
of the binding energy. At this point, we calculate the s.p. wave
functions for all the states below the Fermi surface, the hole states,
and those for a large number of states above it, the particle
states. This calculation is done by assuming that the system is
confined in a spherical box with infinite walls.

As it is traditionally done \cite{rin80}, in this DRPA approach the
secular equations written in matrix form are solved by diagonalizing
the RPA matrix. The dimensions of this matrix are given by the number
of the p-h pairs contributing to the excitation. Therefore, the
results of these DRPA calculations depend on both the size of the box
considered for the HF calculations and the number of particle states
composing the configuration space. The choice of the values of these
two parameters has been done by controlling the centroid energies of
the giant dipole responses.  We have chosen a box radius of 30 fm for
the calculations in oxygen, calcium and nickel isotopes, of 40 fm for
zirconium and tin isotopes, and of 45 fm for lead. For all the nuclei,
but $^{208}$Pb, we have selected a configuration space such as the
maximum p-h excitation energy is 100 MeV. For $^{208}$Pb this would
have generated matrices with too large dimensions, therefore we have
chosen a maximum value of 60 MeV.  We have taken care that the
centroid energies did not change by more than 0.5 MeV when either the
box size, or the configuration space were enlarged.

  In our self-consistent HF plus RPA approach, the spurious center of
  mass motion, showing up as an isoscalar 1$^-$ excitation,
  should appear at zero energy \cite{row70}. This is what happens in
  the CRPA calculations where we do not observe any effect due to the
  presence of this spurious excited state. On the contrary, this
  spurious state is present in all our DRPA results. This is due to
  the truncation of the configuration space. In the great majority of
  the cases we have investigated this state is easily identifiable and
  is well isolated form the other ones.  On the other hand, we observe
  that for the $^{28}$O, $^{48}$Ni $^{100}$Sn and $^{208}$Pb nuclei
  the spurious state appearing in the range 1.5--3.0 MeV is mixed with
  other excited states. In the DRPA results which will be presented in
  the next section, the spurious states have been eliminated by hand.
  In the study of the isoscalar excitations, and of the transition
  densities, we subtracted the contribution of the spurious state by
  using the prescriptions of Refs. \cite{van81} and \cite{roc12} and
  we found that, from the quantitative point of view, the two
  procedures are equivalent. In
  any case, a comparison with the transition densities obtained
  without the subtraction of the spurious states does not show
  numerically relevant differences.

\section{Results and discussion}
\label{sec:resu}

\subsection{Comparison between CRPA and DRPA}
\label{sec:cdrpa}

The nuclei we have considered are listed in Table \ref{tab:nuclei}.
As already pointed out, in these nuclei, the s.p. levels are fully
occupied, and this implies a spherical shape. Furthermore, the pairing
effects are negligible. These features are
confirmed by the deformed
Hartree-Fock-Bogolioubov calculations of Refs. \cite{hil07,del10}.

\begin{table}[ht]
\begin{center}
\begin{tabular}{rcccccc}
\hline \hline
    &~~& $\omega_{\rm sep}$  &~~& $\omega_{\rm PDR}$ &~& $\omega_{\rm GDR}$ \\
\hline
$^{  16}$O  &&  15.5  && 13.71  && 25.11 \\
$^{  22}$O  &&  15.0  &&  8.94  && 24.22 \\
$^{  24}$O  &&  13.5  &&  6.85  && 21.42 \\
$^{  28}$O  &&  12.0  &&  7.32  && 19.32 \\ \hline
$^{  40}$Ca  &&  13.0 &&  12.49 && 21.16 \\
$^{  48}$Ca  &&  12.0 &&  11.12 && 21.72 \\
$^{  52}$Ca  &&  11.0 &&   8.73 && 18.61 \\
$^{  60}$Ca  &&  11.0 &&   9.07 && 18.54 \\ \hline
$^{  48}$Ni  &&  12.0 &&  10.52 && 19.06 \\
$^{  56}$Ni  &&  13.5 &&  13.08 && 18.95 \\
$^{  68}$Ni  &&  12.0 &&  10.93 && 18.05 \\
$^{  78}$Ni  &&  11.5 &&  10.60 && 18.90 \\ \hline
$^{  90}$Zr  &&  12.5 &&  11.60 && 17.61 \\ \hline
$^{ 100}$Sn  &&  11.0 &&  10.74 && 17.12 \\
$^{ 114}$Sn  &&  11.0 &&  10.16 && 19.32 \\
$^{ 116}$Sn  &&  12.0 &&   8.90 && 16.47 \\
$^{ 132}$Sn  &&  11.0 &&   9.01 && 16.50 \\ \hline
$^{ 208}$Pb  &&  10.0 && 8.34 && 14.47 \\
\hline \hline
\end{tabular}
\end{center}
\caption{\small Nuclei considered in our calculations.
For each nucleus we show the energy used to separate 
the PDR and GDR regions, $\omega_{\rm sep}$, and 
the energies $\omega_{\rm PDR}$ and $\omega_{\rm GDR}$ 
of the  discrete states we consider representative of these two regions.
All the energy values are expressed in MeV.}
\label{tab:nuclei}
\end{table}

In Ref. \cite{co12a}, we used the oxygen, calcium, nickel and tin
isotopes indicated in Table \ref{tab:nuclei} to compare the results
obtained with various microscopic mean-field models describing their
ground states. We found a remarkable agreement between the results
obtained with our HF approach and those generated by Skyrme HF and
relativistic Hartree calculations.

\begin{figure}[ht]
\begin{center}
\includegraphics[scale=0.4]{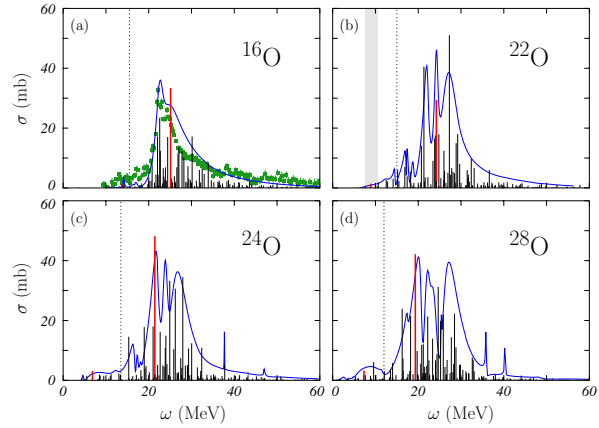} 
\caption{\small (Color on line)
Total photoabsorption cross sections calculated with CRPA (continuous
blue lines), and DRPA (solid vertical lines). The dotted vertical lines
indicate the \epy\/ energies chosen to separate the regions of
the pygmy and giant strengths. We have indicated in red the
discrete levels corresponding to the $\omega_{\rm PDR}$ and $\omega_{\rm GDR}$ 
energies. The total photoabsorption data in the $^{16}$O panel have been taken from
Ref. \cite{ahr75}. The gray region in the $^{22}$O panel indicates the
empirical PDR region found in Ref. \cite{lei01}. 
  }
\label{fig:photo_oxy}
\end{center}
\end{figure}
\begin{figure}[hb]
\begin{center}
\includegraphics[scale=0.4]{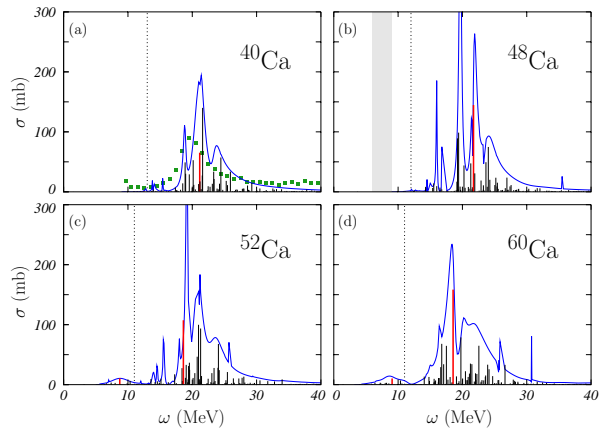} 
\caption{\small  (Color on line)
The same as Fig. \ref{fig:photo_oxy}, for the calcium
  isotopes we have considered. The data in the $^{40}$Ca panel are
  from Ref. \cite{ahr75}. The gray area in the $^{48}$Ca panel indicates the PDR region
  identified in Ref. \cite{har04}.}
\label{fig:photo_ca}
\end{center}
\end{figure}
\begin{figure}[ht]
\begin{center}
\includegraphics[scale=0.4]{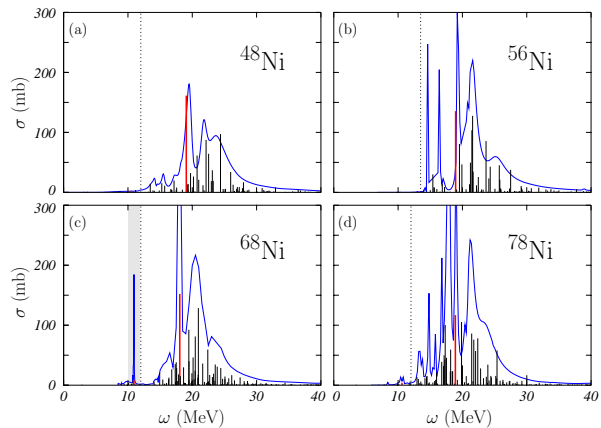} 
\caption{\small  (Color on line)
 The same as Fig. \ref{fig:photo_oxy}, for the nickel 
  isotopes we have considered. The gray area in the $^{68}$Ni panel
  indicates the PDR identified in Ref. \cite{wie09}.
  }
\label{fig:photo_ni}
\end{center}
\end{figure}
\begin{figure}[hb]
\begin{center}
\includegraphics[scale=0.4]{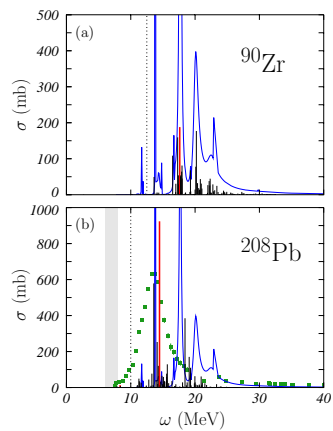} 
\caption{\small  (Color on line)
  The same as Fig. \ref{fig:photo_oxy}, for the
  $^{90}$Zr and the  $^{208}$Pb nuclei. In the $^{208}$Pb panel, the
  experimental data  are from Ref. \cite{ahr75} and the gray area indicates
  the PDR region identified in Ref. \cite{rye02}.
  }
\label{fig:photo_zr}
\end{center}
\end{figure}
\begin{figure}[ht]
\begin{center}
\includegraphics[scale=0.4]{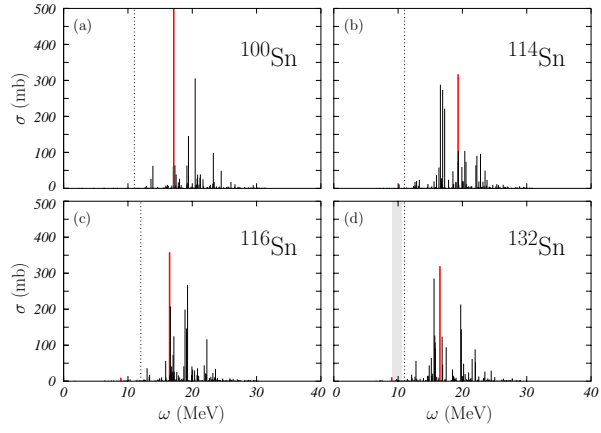} 
\caption{\small  (Color on line)
The same as Fig. \ref{fig:photo_oxy}, for the tin 
  isotopes we have considered. The gray area  in the $^{132}$Sn panel indicates the PDR region
  identified in Ref. \cite{adr05}.
  }
\label{fig:photo_sn}
\end{center}
\end{figure}
\begin{figure}[hb]
\begin{center}
\includegraphics[scale=0.5]{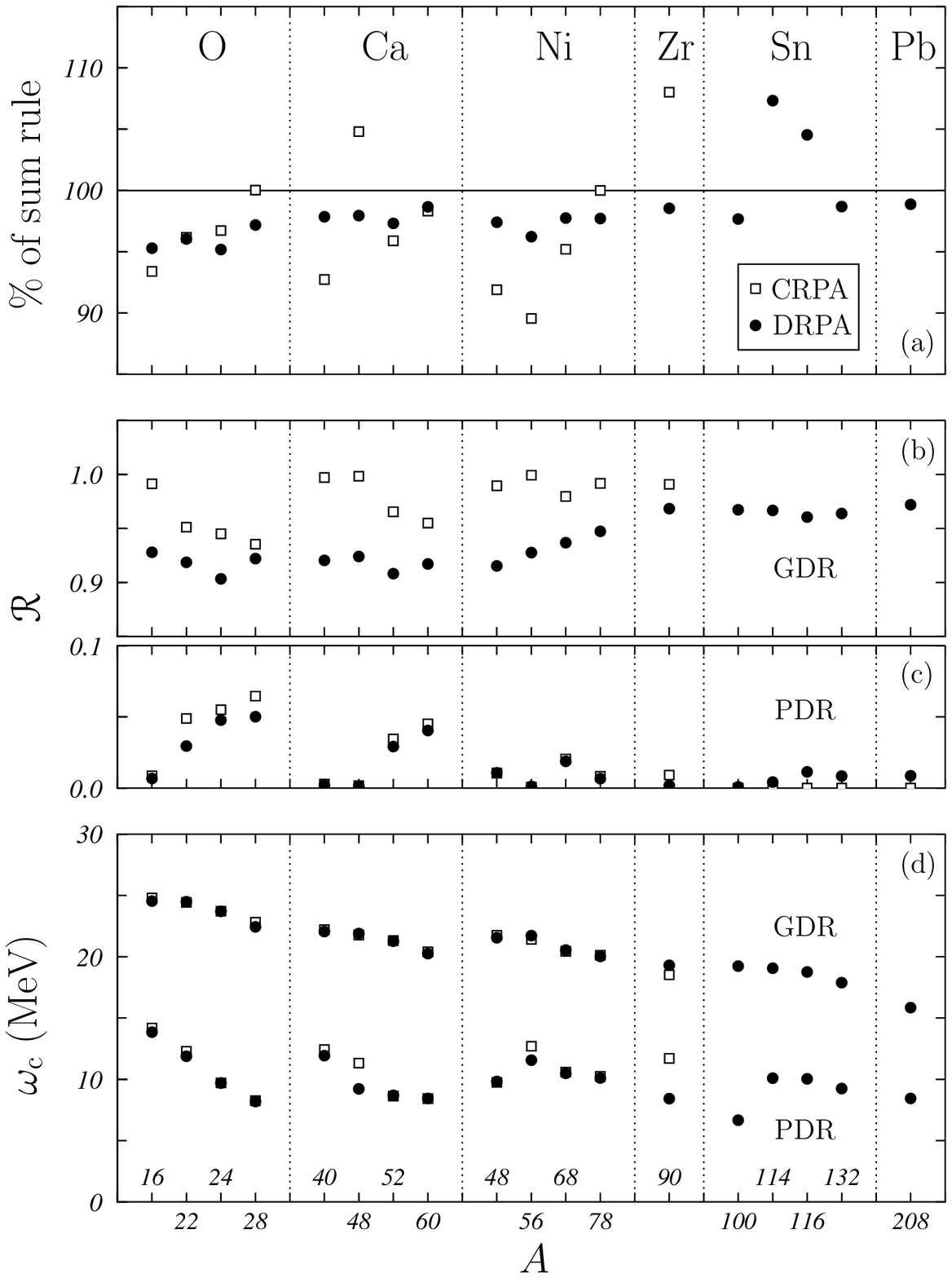} 
\caption{\small Comparison between CRPA (open squares) and DRPA (full circles) results. 
In the panel (a) we show the ratio between the
  energy weighted sum rule obtained by integrating the cross sections
  of Figs. \ref{fig:photo_oxy}-\ref{fig:photo_sn} and the values obtained by considering the TRK
  sum rule plus an estimated 0.5 enhancement factor. In the panels (b)
  and (c) we compare the ${\cal R}$ values defined in Eq. (\ref{eq:R})
  calculated for the GDR and the PDR regions, respectively. In
  panel (d) we show the GDR and PDR centroid energies as given by Eq. 
  (\ref{eq:wc}). }
\label{fig:cd}
\end{center}
\end{figure}

We show in Figs. \ref{fig:photo_oxy}-\ref{fig:photo_sn} the total photoabsorption cross sections obtained in our calculations
of the dipole responses.  In these figures the
vertical bars show the results obtained with DRPA. For all the
isotopes lighter than tin, we show with solid blue lines the cross
section obtained with CRPA. As explained above, for the four tin isotopes (Fig. \ref{fig:photo_sn}) and
also for $^{208}$Pb (Fig. \ref{fig:photo_zr}b) we have DRPA results
only.

For $^{16}$O, $^{40}$Ca and $^{208}$Pb nuclei
we compare the results of our calculations 
with the experimental data of Ref. \cite{ahr75}. This
comparison emphasizes the well known merits and faults of the
RPA calculations in the description of the giant resonances 
\cite{spe91,bor98,har01}. The presence of a peak generated
by the collective excitation of the nucleus is predicted by our
calculations. The position of the main peaks is reasonably well
reproduced. The CRPA centroid energies for $^{16}$O and $^{40}$Ca are
24.8 and 22.0 MeV to be compared with the experimental values of 22.9
and 20.0 MeV respectively. In $^{208}$Pb the DRPA
centroid energy is 15.84 MeV and the experimental value is 14.7
MeV. Since in these calculations the position of the GDR is dominated
by the isospin dependent part of the interaction \cite{deh82,co85},
these discrepancies may indicate that this term of the D1M interaction
is too strong.

The RPA calculations satisfy the energy weighted sum rule, which can
be calculated by integrating the cross sections presented in the
figures. For the $^{16}$O, $^{40}$Ca and $^{208}$Pb nuclei we obtained
the values of 336, 834 and 4420 mb MeV, respectively. These values
should be compared with those obtained by using the Thomas-Reiche-Khun
(TRK) sum rule corrected by the enhancement factor 
 related to the
  isospin-dependent, and eventually tensor, terms, of 
  the interaction used.  The enhancement
  factors 
  should be calculated for each nucleus considered. For the purposes
  of the present paper we used a simple estimation 
  based on the values published in Ref. \cite{tra87}.  
  In the case of the D1M interaction we have used a value of 0.51.
The TRK values corrected for 
this enhancement factor are 360, 900, and 4470 mb MeV for
$^{16}$O, $^{40}$Ca and $^{208}$Pb, respectively, indicating a
reasonable agreement with the CRPA values. 

The CRPA calculations consider only the escape width of the resonance
and do not take into account the spreading width. As a consequence of
this fact we observe that the strength is too concentrated in the
resonance region. This drawback becomes more important the heavier is
the nucleus when the second type of width dominates. We observe a
reasonable agreement in $^{16}$O, but in $^{40}$Ca the
width of the resonance is not correctly reproduced. This feature has
been discussed in more detail in Ref. \cite{don11a}. 

We want to discuss the reliability of the DRPA results as compared to
the CRPA ones. This comparison is important to have confidence on the
DRPA results in lead and tin isotopes where we could not perform CRPA
calculations. The results shown in
Figs. \ref{fig:photo_oxy}-\ref{fig:photo_sn} indicate a good agreement
between the position of the peaks in the two types of
calculations. 

We have calculated the sum rule exhaustion of both types of
calculations for all the nuclei, and we present them in the panel (a)
of Fig. \ref{fig:cd}. The important information given in this
  figure is the comparison between results of the CRPA, 
 open squares, and those obtained in DRPA calculations, full circles. 
To simplify the presentation, these values are divided by the
TRK values corrected with the unique 0.51  enhancement factor.  
The scale of the figure emphasizes the
  differences between CRPA and DRPA results, but 
we observe that most of our results are within a range of 5\% around
the expected theoretical values. 
The largest deviations between CRPA and
  DRPA results are found in the double-closed shell nuclei 
$^{40}$Ca, $^{40}$Ca, $^{48}$Ni, $^{56}$Ni and $^{90}$Zr.  Also in these
cases, however, the differences  are at most 10\%.  

  The behaviour of the DRPA values seems more regular than
  that of the CRPA results. Remarkable deviations from this
  behaviour are shown by the $^{114}$Sn and $^{116}$Sn nuclei. It is
  possible that, in these cases, shell effects are important
  in the evaluation of the enhancement factors.

We further carried on the comparison between the CRPA and DRPA results
by separately analyzing the pygmy and the giant resonance regions. For
this reason, we divide the dipole response in two regions
representative of the PDR and GDR.  The values of the energies that
separate these two regions, $\omega_{\rm sep}$, are given in Table
\ref{tab:nuclei} for all the isotopes studied, and are indicated by
the vertical dotted lines in
Figs. \ref{fig:photo_oxy}-\ref{fig:photo_sn}.  We checked that the
values of the quantity
\begin{equation}
{\cal R} \, = \,
\frac{ \displaystyle \int_{\omega_i}^{\omega_f} B(E1;\omega) \, {\rm d}\omega 
}{ \displaystyle \,       \int_0^\infty B(E1;\omega)\, {\rm d}\omega }
\, ,
\label{eq:R}
\end{equation}
calculated with CRPA and DRPA were approximately the same in each of 
the two regions, 
obviously for all those isotopes where the CRPA calculations have
been done. In our approach, the $B(E1;\omega)$ value is calculated as
\beq
B(E1;\omega)\, = \, \left| \displaystyle \sum_{ph=1}^{N_{ph}} \, 
b_{ph}(E1;\omega) \right|^2 \, ,
\eeq
where the relation with the usual $X$ and $Y$ RPA amplitudes is given by
\beq
b_{ph}(E1;\omega) \, = \, e_{\rm eff}
\left[ X_{ph}(\omega) - Y_{ph}(\omega) \right] \, 
\int {\rm d}r \, r^3 \, \rho_{ph}(E1;\omega,r) 
\,\,\,.
\eeq
In the above equation, we used the common definition of the effective
charge for the electric dipole excitation  
\beq
e_{\rm eff} = \left\{
\begin{array}{rl} \displaystyle
\frac{N}{A} \,e   & \mbox{for protons} \\   
   & \\
 - \displaystyle \frac {Z} {A} \,e & \mbox{for neutrons \,\,.}  
\end{array}
\right.
\eeq
The p-h transition density is defined as 
\beq
\rho_{ph}(E1;\omega,r) \, = \, \sqrt{\frac{3}{4 \pi}} \, (-1)^{j_p +\half} 
\, \sqrt{(2j_p+1)(2j_h+1)} \, 
\,\threej {j_p}{1}{j_h}{\half}{0}{-\half} 
\,R_p(r) \, R_h(r) \, ,
\label{eq:rhoph}
\eeq
where $R(r)$ indicates the radial part of the s.p. wave
function, $j$ is the total angular momentum and we have used the
traditional symbol to indicate the Wigner 3-j coefficient.
The above expression has been
obtained by using the $|l \half j \rangle$ angular momentum coupling. 

In Eq. (\ref{eq:R}) we used $\omega_i=0$ and $\omega_f=\omega_{\rm
  sep}$ for the PDR, while for the GDR region we used
$\omega_i=\omega_{\rm sep}$ and $\omega_f=$40 MeV for the oxygen
isotopes and $\omega_f=$30 MeV for all the other nuclei.  The results
of our calculations are shown in the panels (b) and (c) of
Fig. \ref{fig:cd} for GDR and PDR, respectively. The values obtained
with CRPA are indicated by the open squares, and those obtained with DRPA
by the full circles. We remark a good agreement between these results. 

We further tested the reliability of the DRPA results as compared to
the CRPA ones, and also the validity of our choices in selecting
the $\omega_{\rm sep}$ values, by calculating the centroid energies of
the two regions, with the usual expression
\begin{equation}
\omega_c \, = \, 
 \frac{\displaystyle \int_{\omega_i}^{\omega_f} \omega \, B(E1;\omega) \, {\rm d}\omega } 
 {    \displaystyle   \int_{\omega_i}^{\omega_f} B(E1;\omega) \, {\rm d}\omega } 
\,\, .
\label{eq:wc}
\end{equation}
 
The values obtained are shown in the panel (d) of
Fig.~\ref{fig:cd}. We find, again, a good agreement between the
results of the two calculations.

In the DRPA results, for each isotope, we have selected two states which are representative of the main characteristics of the PDR
and GDR regions. In Figs. \ref{fig:photo_oxy}-\ref{fig:photo_sn} these
states are identified by the red vertical lines and the corresponding
excitation energies are indicated in Table \ref{tab:nuclei} as
$\omega_{\rm PDR}$ and $\omega_{\rm GDR}$.

\subsection{Photoabsorption cross sections}

The photoabsorption cross sections for the four oxygen isotopes we
have selected are shown in Fig. \ref{fig:photo_oxy}. In $^{16}$O the
number of protons and neutrons is the same. When the number of
neutrons increases, also the strength below $\omega_{\rm sep}$
increases.  This is also evident in the panel (c) of
Fig. \ref{fig:cd}, where it can be seen how the ratio ${\cal R}$,
calculated for the PDR region, increases with the number of neutrons
for the oxygen isotopes. In any case, the maximum value of ${\cal R}$
is around 5\% of the total strength.

The gray area in panel (b) of Fig. \ref{fig:photo_oxy} indicates the
region empirically identified as PDR in Ref. \cite{lei01}. We observe
that the value of $\omega_{\rm PDR}$ is compatible with these
experimental findings.

We show in Fig. \ref{fig:photo_ca}, the results for the calcium
isotopes.  Also in this case, we observe that the presence of dipole
strength in the low energy region, below $\omega_{\rm PDR}$, increases
together with the neutron number. This is evident by observing the
behavior of ${\cal R}$  in the panel (c) of
Fig. \ref{fig:cd} which increases up to a maximum value of about 5\%
for $^{60}$Ca.

The PDR region empirically identified in Ref. \cite{har04} in the
$^{48}$Ca (see panel (b) of Fig. \ref{fig:photo_ca}) is lying at
energies smaller than those predicted by our model. We remark,
however, that we do not observe much PDR strength in $^{48}$Ca. The
corresponding value of ${\cal R}$ in the panel (c) of
Fig. \ref{fig:cd} is smaller than 1\%.  As we mentioned in the
Introduction, we have carried out the same calculation with the D1S
interaction, and obtained similar results. This indicates that the
discrepancy with the observed PDR strength should not be attributed to
some specific features of the interaction.  The most probable source of
this discrepancy is related to the intrinsic limitations of the RPA
calculations, specifically to the fact that only one particle-one hole
(1p1h) configurations are considered. In fact, the RPA results shown
in Ref. \cite{har04}, are similar to ours, while some dipole strength
appears at low energies when configurations beyond 1p1h were
considered within the framework of the extended theory of finite Fermi
systems \cite{kam04}. The relevance of effects beyond RPA in the PDR
region has been confirmed by the self-consistent second RPA results of
Ref. \cite{gam11}, obtained with a Skyrme interaction.

We show in Fig.~\ref{fig:photo_ni} the results for the nickel
isotopes. Also in this case, we observe features analogous to those
found in the two previous figures. We should remark that $^{48}$Ni has
a neutron deficit, and $^{56}$Ni is the isotope with the same number
of protons and neutrons. In the latter isotope the PDR strength is
almost negligible.

The energy range of the empirical PDR in $^{68}$Ni \cite{wie09}
includes the very sharp peak at 10.93 MeV, which identifies the pygmy
region.  In the case of the nickel isotopes, the relation between the
appearance of a PDR and the increase of the neutron number is not any
more as straightforward as in the previous cases as it is shown in the
panel (c) of Fig. \ref{fig:cd}. The shell structure of each individual
isotope plays a remarkable role.

We show in Fig.~\ref{fig:photo_zr} the results for $^{90}$Zr and
$^{208}$Pb.  The nucleus $^{90}$Zr is the heaviest isotope where we
could perform our CRPA calculations without losing numerical
accuracy.  We observe again a good agreement with the DRPA
results. There is dipole strength below $\omega_{\rm sep}$ carrying
slightly less than 2\% of the total strength. 

As already mentioned before, for the $^{208}$Pb nucleus we performed
only DRPA calculations. We have already commented that in this nucleus
the spreading width is more important than the escape width, this
latter one well described by the CRPA theory. 
In analogy to what we found  for the $^{48}$Ca nucleus,
the pygmy strength
empirically identified \cite{rye02} is lying in a region where our
calculations do not predict relevant dipole strength. 
On the contrary, self-consistent RPA calculations where Skyrme
  interactions are used find remarkable
  strength in this region \cite{roc12,lan11}.
  In $^{208}$Pb the excitations beyond RPA are even more important than
in $^{48}$Ca.

We conclude this general view of the RPA dipole strengths by
discussing the results obtained for the tin isotopes and shown in
Fig.~\ref{fig:photo_sn}. Our calculations have been done with DRPA
only. Also in this case, we observe an increase of the pygmy strength
with increasing neutron number. However, the relative contribution of
the PDR with respect to the global strength is of the order of 1-2\%,
as can be seen in panel (c) of Fig. \ref{fig:cd}. The empirical PDR
region identified in the $^{132}$Sn nucleus \cite{adr05} includes some
of the strength found in our DRPA calculations.

\subsection{Collectivity of the resonances}

One of the main questions concerning the structure of the PDR is up to
which extent it can be considered as a collective excitation of the
nucleus. The identification of the parameters which can define the
degree of collectivity is not an easy task. In a RPA approach one
should consider parameters related to the number of the p-h pairs
participating to the excitation, and also to the degree of coherence
of the excitation. 

In DRPA, the $X$ and $Y$ amplitudes of a given
excited state at energy $\omega$ must verify the normalization
relation 
\beq
\sum_{ph=1}^{N_{ph}} \left[X_{ph}^2(\omega) - Y_{ph}^2(\omega) \right] =1 
\, ,
\label{eq:norm1}
\eeq
where $N_{ph}$ indicates the total number of p-h excitations. 
To estimate the collectivity of a DRPA excited state
we used the ratio \cite{co09b}
\beq
\doc(\omega) = \frac{N^*} {N_{ph} } 
\, ,
\label{eq:doc}
\eeq
where \nst is the number of p-h configurations such as
\beq
\left[X_{ph}^2(\omega) - Y_{ph}^2(\omega) \right] \, \ge \,  \frac{1}{N_{ph}}
\label{eq:nstar}
\, .
\eeq
The two extreme values of $\doc$ are 1 in the fully collective case,
and $1/N_{ph}$ when the excitation is produced by a single p-h
pair. 

To study the relative contribution of protons and neutrons to the
  excitation, we have also considered the index
\beq
{\cal{N}}(\omega) \, = \, \sum_{ph=1}^{N_{ph}} \, \delta_{ph,\nu} \, \left[X_{ph}^2(\omega) - Y_{ph}^2(\omega) \right] \, ,
\eeq
which indicates the contribution of the neutron p-h pairs to the
normalization given in Eq.~(\ref{eq:norm1}).

These two indexes are related to the relevant number of p-h pairs in the
excitation of the resonance. In order to clarify the degree of
collectivity it is important to consider also the coherence of the
excitation, as pointed out in Ref. \cite{lan09}. To this purpose we
have defined the index
\beq
\coh(\omega) \, = \, B(E1;\omega)
\left(\sum_{ph=1}^{N_{ph}} \, \left| b_{ph}(E1;\omega)\right| \right)^{-2}
 \, \, .
\label{eq:coh}
\eeq 
The second term of the above expression calculates the value of
$B(E1;\omega)$ that would be obtained if all the contributions of the
various p-h pairs would add coherently. The maximum, and global,
coherence of the RPA p-h pairs provides $\coh=1$.

\begin{figure}[hb]
\begin{center}
\includegraphics[scale=0.5]{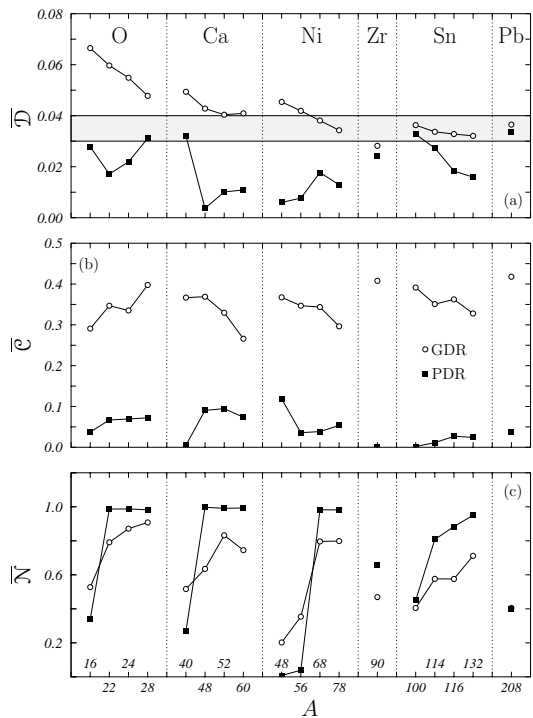} 
\caption{\small Collectivity indexes for the giant and pygmy
  regions. In each panel, the open circles indicate the results
  related to the GDR and the full squares those related to the PDR. 
  We show in (a) the values of $\overline{\doc}$, in (b) those of $\overline{\coh}$, and in (c) 
  those of $\overline{\cal{N}}$.
  The gray band in panel (a) represents the range
  of values of $\doc$ obtained for the low-lying 3$^-$ states of
  $^{16}$O, $^{40}$Ca, $^{132}$Sn and $^{208}$Pb. The values of $\coh$
  we obtained for these states are above 0.9.
  The lines have been drawn to guide the eyes. 
  }
\label{fig:indexes}
\end{center}
\end{figure}

In the present study, we have averaged the indexes 
 ${\doc}$, ${\cal{N}}$, and ${\coh}$
in both the
pygmy and giant regions by using the expression:
\beq
\overline{\cal{J}}\, = \, \displaystyle 
\frac{\displaystyle\sum_{\omega_i}^{\omega_f} \, B(E1;\omega)\, {\cal{J}}(\omega)} 
{\displaystyle \sum_{\omega_i}^{\omega_f} \, B(E1;\omega)} 
\label{eq:ave}
\, , 
\eeq 
where, for each region, $\omega_i$ and $\omega_f$ have been
defined in Sect. \ref{sec:cdrpa} and the values are given in Table
\ref{tab:nuclei}.  In Fig. \ref{fig:indexes}
we show the indexes $\overline{\doc}$, $\overline{\cal{N}}$, and
$\overline{\coh}$ calculated for both the giant (open circles) and
pygmy (solid squares) regions.  In the figure, the lines joining the
symbols have been drawn to guide the eyes.

The values of $\overline{\doc}$ depend on the dimension of the
s.p. configuration space, therefore a comparison among the different
nuclei is not very meaningful. On the other hand, the comparison of
the results obtained in the same nucleus is relevant.  The important
result is that, as we observe in panel (a) of Fig. \ref{fig:indexes},
the values of $\overline{\doc}$ for the GDR are always larger than
those obtained in the pygmy region.

To have an indication of the degree of collectivity we have calculated
the value of $\doc(\omega)$, Eq. (\ref{eq:doc}), for the collective
low-lying $3^-$ states of the $^{16}$O, $^{40}$Ca, $^{132}$Sn and
$^{208}$Pb nuclei. The gray band gives the upper and lower bounds of
these values.  While the values of $\overline{\doc}$ for the GDRs are
always larger, or compatible, with the $3^-$ band, those of the PDRs
are usually smaller, with the exceptions of the $^{28}$O, $^{40}$Ca,
$^{100}$Sn and $^{208}$Pb, where they are inside the band.

We show in the panel (b) of Fig. \ref{fig:indexes} 
the values we obtained for 
$\overline{\coh}$. For the  $3^-$ states mentioned above, the values
of $\coh$ are larger than 0.9, well above than 
the values obtained for the GDR that, on the other hand, are always 
larger than
those related to the PDR. The combined results of\/
$\overline{\doc}$ and $\overline{\coh}$ indicate that the PDRs are less
collective than the GDRs. 
 
We show in the panel (c) of Fig. \ref{fig:indexes} the values of
$\overline{\cal{N}}$ for the PDR and GDR regions.  In the GDR case
(see open circles), we found the values of\/ $\overline{\cal{N}}$  
between 0.2 and 0.8, with the only exceptions of the $^{24}$O and
$^{28}$O nuclei which have larger values.  
This fact indicates that, in general, both
protons and neutrons contribute to the excitation of the GDR. Each
nucleus has its own peculiarities according to the closure of the
neutron shells.  In general, in nuclei with neutrons in excess, the neutron
contribution is larger than that of the protons.  The situation
changes for the PDR (see full squares). In this case, all the nuclei
with neutron excess, except $^{90}$Zr and $^{208}$Pb, 
show values of\/ $\overline{\cal{N}}$ close to
1.  We observe few cases where the $\overline{\cal{N}}$ values for
the GDR are larger than for the PDR. This happens for the $N=Z$ nuclei 
$^{16}$O, $^{40}$Ca, $^{56}$Ni and $^{100}$Sn, and for the $Z>N$ nucleus $^{48}$Ni.

\subsection{Transition densities}

The DRPA calculation allows a detailed study of the p-h structure of each
excited state. We have exploited this feature by calculating the dipole
transition density 

\begin{figure}[hb]
\begin{center}
\includegraphics[scale=0.4]{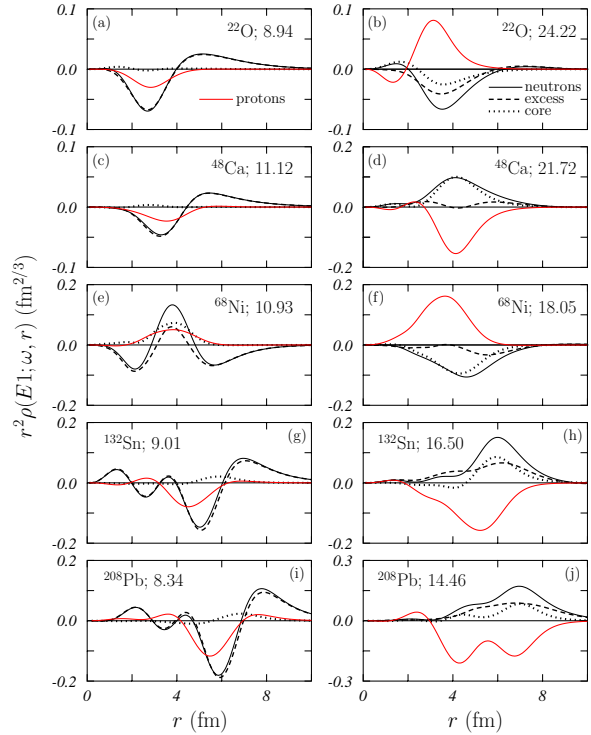}
\caption{\small (Color on line) 
  Transition densities, as given by Eq. (\ref{eq:trans1}), for protons (full red
  lines) and neutrons, full black lines. 
  The black dotted curves show the contributions of the neutrons of
  the core, while those of the neutrons in excess are shown by the black
  dashed lines. The number in each panel indicates the 
  excitation energy $\omega_{\rm PDR}$ (left panels) or $\omega_{\rm
    GDR}$ (right panels), in MeV, of the specific states studied.    
  }
\label{fig:trans}
\end{center}
\end{figure}

\beqn
\rho(E1;\omega,r) \,= \, \sum_{ph=1}^{N_{ph}} \,
\left[ X_{ph}(\omega) - Y_{ph}(\omega) \right] 
\, \rho_{ph}(E1;\omega,r) \, ,
\label{eq:trans1}
\eeqn
where $\rho_{ph}(E1;\omega,r)$ is given in Eq. (\ref{eq:rhoph}). The
calculations have been done for those states that were selected to
characterize the pygmy and giant regions. The excitation energies of
these states, $\omega_{\rm PDR}$ and $\omega_{\rm GDR}$, are listed in
Table \ref{tab:nuclei}, and the related photoabsorption cross sections
are indicated by the red vertical lines in
Figs.~\ref{fig:photo_oxy}-\ref{fig:photo_sn}.

In Fig. \ref{fig:trans} we present the dipole transition densities,
multiplied by $r^2$, for those nuclei with an empirical indication of
PDR. Similar results have been found for all the other nuclei we have
considered. In the figure, the solid red and black lines have been
obtained by considering only the contribution to the p-h sum in
Eq. (\ref{eq:trans1}) of the protons and neutrons, respectively.  The
neutron transition densities have been further analyzed by separating
the contribution of the neutrons of the core, represented by the
dotted black lines, and that of the neutrons in excess, indicated by the
dashed black lines.  The numbers in each panel indicate, in MeV, the
excitation energy. The states in the left column belong to the PDR
region, those in the right column to the GDR one.

All the results we have obtained present common trends. In the PDR
states proton and neutron transition densities are in phase, while
they are out of phase in the GDR states. This statement is clear in
the lighter nuclei, $^{22}$O and $^{48}$Ca. In heavier nuclei the
number of the nodes of the two transition densities is different,
therefore the situation is more involved. For the PDR it is evident,
however, that at the nuclear surface both transition densities are
always in phase. The transition
densities related to the GDR do not show this kind of ambiguities. In
these cases proton and neutron transition densities are evidently
always out of phase.

There is another feature common to all the results presented in
Fig. \ref{fig:trans}. In the transition densities of the GDR, the
contribution of the neutrons in excess is comparable to that of the
neutrons of the core. On the contrary, the PDR transition densities
are dominated by the neutrons in excess.

\begin{figure}[hb]
\begin{center}
  \includegraphics[scale=0.4]{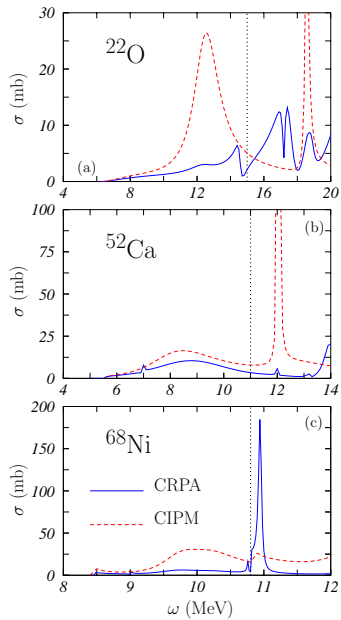}
\caption{\small  (Color on line)
Total photoabsorption cross sections obtained with 
  CRPA calculations (full blue lines) and by switching off the
  residual interaction, CIPM (red dashed lines). The black dotted vertical
  lines indicate the $\omega_{\rm sep}$ energies of Table \ref{tab:nuclei}. 
  }
\label{fig:hf}
\end{center}
\end{figure}

\subsection{Role of the residual interaction}

The results we presented so far indicate that the structures of the
pygmy and of the giant dipole excitations are different.  
We studied 
whether the different structures imply
different sensitivities to the various terms of the effective
nucleon-nucleon force.  
In Fig. \ref{fig:hf}, for
three nuclei where we have clearly identified the 
presence of pygmy dipole
strength, we compare the CRPA results with those obtained by completely
switching off the residual interaction. We call continuum independent
particle model (CIPM) this last type of calculations.

As we have done in Figs. \ref{fig:photo_oxy}-\ref{fig:photo_sn}, we
indicate the separation between the PDR and GDR regions with a dotted
vertical line located at $\omega_{\rm sep}$ energies. In all the cases
presented in the figure the CIPM cross sections are larger than the
CRPA ones in the PDR region. We observe the
largest difference in the $^{22}$O nucleus, where the large and wide
peak of the CIPM result is due to the $(f_{7/2})(1d_{5/2})^{-1}$
neutron transition. In the RPA calculation the residual interaction
moves the main contribution of this p-h pair in the GDR region.
The results for the $^{52}$Ca and $^{68}$Ni nuclei are less extreme.

These results indicate a remarkable sensitivity of the dipole
strength 
to the presence of the residual
interaction in the RPA calculations. 
To identify the components of 
the interaction to which the PDR is more sensitive, we adopted the
strategy to switch on and off the various terms of the interaction.  
Unfortunately, 
when a finite 
range interaction is used, each term of the interaction
contributes to the other channels via exchange terms. 
We simplify our study by constructing 
a zero range interaction to be used by considering  
only direct RPA matrix elements. In this case,  
we can completely switch off and on 
the contributions of the 
individual terms of the force and observe their effect on the RPA
results. We constructed a
zero-range interaction of Landau-Migdal (LM) type:
\beq v_{\rm  LM}(\br_1-\br_2) 
\, =\,  \left[ F + G \, \bsigma(1) \cdot \bsigma(2) + F' 
\,  \btau(1) \cdot \btau(2) + G' \, \bsigma(1) \cdot \bsigma(2) \btau(1)
  \cdot \btau(2) \right] \delta(\br_1 - \br_2)
\label{eq:LM} 
\eeq
where $F$, $F'$, $G$ and $G'$ are constants whose values have been
chosen to reproduce, approximately, the DRPA responses obtained with
the D1M Gogny interaction.

We discuss here only the results we have obtained for the $^{68}$Ni
case. We show in Table \ref{tab:omega} the centroid energies,
$\omega_c$, for the pygmy and giant regions obtained in DRPA
calculations with various interactions. As we can see, the values
obtained with the complete LM interaction reproduce rather well the
centroid energies found with the D1M force.

By setting to zero the $G$ and $G'$ constants, we have verified that
the electric dipole response is essentially insensitive to the spin
dependent terms of the interaction. We present in Table
\ref{tab:omega} the results obtained by setting to zero $F$ and $F'$.
The calculations done with $F=0$ do not modify the centroid energy of
the GDR, and increase that of the PDR by only the 2\%. The
calculations with $F'=0$ lower the PDR centroid energy by 3\%, but
reduce the centroid energy of the GDR by 30\%.
We observed similar results in all the cases we have investigated. The
PDR has a small sensitivity to the scalar and isospin channels. The
contribution of the scalar term is attractive, i.e. it lowers the values of
$\omega_c$, while that of the isospin channel is repulsive.

\begin{table}[ht]
\begin{center}
\begin{tabular}{lcccc}
\hline \hline
    &~~& PDR  &~~& GDR \\
\hline
D1M      &&  10.49  && 20.54 \\
LM       &&  10.69  && 19.65 \\
LM ($F=0$)  &&  10.91  && 19.64 \\
LM ($F'=0$) &&  10.37  && 13.67 \\
\hline \hline
\end{tabular}
\end{center}
\caption{\small Centroid energies, $\omega_c$, in MeV, of the PDR and
  GDR in $^{68}$Ni obtained with different types of calculations. The
  D1M results indicate the values obtained in the HF+DRPA calculation
  with the D1M interaction. The values listed in the LM row show the
  results obtained by using, in DRPA calculations, the $v_{\rm  LM}$
  interaction of Eq. (\ref{eq:LM}) with $F=-100$, $F'=400$, $G=200$ 
  and $G'=300$ MeV fm$^3$. The last two rows show the values obtained
  by setting $F=0$, and $F'=0$.
}
\label{tab:omega}
\end{table}

\section{Conclusions}
\label{sec:conc}

We have investigated the pygmy and giant dipole excitations in a set
of 18 spherical nuclei in various mass regions. We have conducted our
study by doing self-consistent RPA calculations with the finite range
D1M Gogny interaction. We carried out discrete and continuum RPA
calculations, the latter ones for 13 nuclei lighter than
$^{100}$Sn. We have verified the convergence of the discrete and
continuum results. For each nucleus considered we have separated the
two energy regions identifying the pygmy and the giant responses. 
Centroid energies and sum rule exhaustion are conserved in both
discrete and continuum RPA calculations in each energy region. 

The main results of our investigation are summarized here below.
\begin{enumerate}
\item The dipole strength at energies around the nucleon emission
  threshold increases with the neutron excess. 
\item The PDR exhausts about 5\% of the total energy weighted sum
  rule, while the GDR about the 90\%.
\item The values of the collectivity indexes $\overline{\doc}$ 
      and $\overline{\coh}$, Eqs. (\ref{eq:doc}) and (\ref{eq:coh}),
  indicate that the GDRs are more collective than the PDRs. 
\item The PDR is dominated by the neutron p-h excitations, while in the
  GDR the contributions of both proton and neutron excitations are comparable.
\item At the nuclear surface, proton and neutron transition densities
  are in phase in the PDR region, while they are out of phase in the
  GDR region.
\item The neutron transition densities of the PDR states are dominated
  by the motion of the neutrons in excess. In the GDR both the neutrons 
  in excess and those of the core contribute to the excitation. 
\item The scalar and isospin terms of the effective interaction have
  opposite effects on the PDR, the former is attractive and the latter
  repulsive. In any case these effects are much smaller than those
  produced by the interaction on the GDR.
\end{enumerate}

Our investigation indicates that, 
as it has been suggested in Refs. \cite{rei10,rei12}, 
the emergence of the PDR is an
effect more related to the shell structure of the various isotopes
than to a real collective nuclear motion.

The limits of our investigation are related to the validity of the
intrinsic hypotheses of the theory we used, the RPA.  Specifically, we
refer to the fact that in RPA the excited states are described as
linear combination of one particle-one hole, and one hole-one
particle, excitations. The role of configurations beyond RPA on the
PDR was investigated in \cite{ter07,avd11,gam11} where a
redistribution of the strength was pointed out as the main
effect. This generate strength also at energies lower than those
indicated by RPA calculations \cite{har04}. A detailed comparison with
the future experimental data would require the use of self-consistent
theories beyond RPA \cite{gam12}.  In any case, the main results of
our investigation depend on the global properties of the excitation,
rather than on the detailed distribution of the strength. For this
reason, we believe that they would not be changed by theories more
elaborated than RPA.

\acknowledgments 
This work has been partially supported by the PRIN (Italy) {\sl
  Struttura e dinamica dei nuclei fuori dalla valle di stabilit\`a},
by the Spanish Ministerio de Ciencia e Innovaci\'on under contract
FPA2009-14091-C02-02 and by the Junta de Andaluc\'{\i}a (FQM0220).



\begin{thebibliography}{10}
\expandafter\ifx\csname url\endcsname\relax
  \def\url#1{\texttt{#1}}\fi
\expandafter\ifx\csname urlprefix\endcsname\relax\def\urlprefix{URL }\fi

\bibitem{paa07}
N.~Paar, D.~Vretenar, E.~Khan, G.~Col\`o, Rep. Prog. Phys. 70 (2007) 691.

\bibitem{kre09}
S.~Krewald, J.~Speth, Int. \ Jour. \ Mod. \ Phys. E 18 (2009) 1425.

\bibitem{gor98}
S.~Goriely, Phys. \ Lett. \ B 436 (1998) 10.

\bibitem{dao12}
I.~Daoutidis, S.~Goriely, Phys. \ Rev. \ C 86 (2012) 034328.

\bibitem{arn07}
M.~Arnould, S.~Goriely, K.~Takahashi, Phys. \ Rep. 450 (2007) 97.

\bibitem{boh81}
O.~Bohigas, N.~Van Giai, D.~Vautherin, Phys. \ Lett. \ B 102 (1981) 105.

\bibitem{kri82}
H.~Krivine, C.~Schmit, J.~Treiner, Phys. \ Lett. \ B 112 (1982) 281.

\bibitem{car10}
A.~Carbone, G.~Col\`o, A.~Bracco, L.~G. Cao, P.~F. Bortignon, F.~Camera,
  O.~Wieland, Phys. \ Rev. \ C 81 (2010) 041301.

\bibitem{rei10}
P.~G. Reinhard, W.~Nazarewicz, Phys. \ Rev. \ C 81 (2010) 051303(R).

\bibitem{ina11}
T.~Inakura, T.~Nakatsukasa, K.~Yabana, Phys. \ Rev. \ C 84 (2011) 021302(R).

\bibitem{co09b}
G.~Co', V.~De~Donno, C.~Maieron, M.~Anguiano, A.~M. Lallena, Phys. \ Rev. \ C
  80 (2009) 014308.

\bibitem{ang11}
M.~Anguiano, G.~Co', V.~De~Donno, A.~M. Lallena, Phys. \ Rev. \ C 83 (2011)
  064306.

\bibitem{ang12}
M.~Anguiano, M.~Grasso, G.~Co', V.~De~Donno, A.~M. Lallena, Phys. \ Rev. \ C 86
  (2012) 054302.

\bibitem{co12b}
G.~Co', V.~De~Donno, M.~Anguiano, A.~M. Lallena, Phys. \ Rev. \ C 85 (2012)
  034323.

\bibitem{dec80}
J.~Decharg\`e, D.~Gogny, Phys. \ Rev. \ C 21 (1980) 1568.

\bibitem{gor09}
S.~Goriely, S.~Hilaire, M.~Girod, S.~P\'eru, Phys. \ Rev. \ Lett. 102 (2009)
  242501.

\bibitem{lei01}
A.~Leistenschneider, et~al., Phys. \ Rev. \ Lett. 86 (2001) 5442.

\bibitem{har04}
T.~Hartmann, et. al., Phys. \ Rev. \ Lett. 93 (2004) 192501.

\bibitem{wie09}
O.~Wieland, et~al., Phys. \ Rev. \ Lett. 102 (2009) 092502.

\bibitem{adr05}
P.~Adrich, et. al., Phys. \ Rev. \ Lett. 95 (2005) 132501.

\bibitem{rye02}
N.~Ryezayeva, et. al., Phys. \ Rev. \ Lett. 89 (2002) 272502.

\bibitem{pol12}
I.~Poltoratska, et~al., Phys. \ Rev \ C. 85 (2012) 041304(R).

\bibitem{co98b}
G.~Co', A.~M. Lallena, Nuovo \ Cimento \ A 111 (1998) 527.

\bibitem{bau99}
A.~R. Bautista, G.~Co', A.~M. Lallena, Nuovo \ Cimento \ A 112 (1999) 1117.

\bibitem{don08t}
V.~De~Donno, Nuclear excited states within the random phase approximation
  theory, Ph.D. thesis, Universit\`a del Salento (Italy),
  http://www.fisica.unisalento.it/~gpco/stud.html (2008).

\bibitem{don11a}
V.~De~Donno, G.~Co', M.~Anguiano, A.~M. Lallena, Phys. \ Rev. \ C 83 (2011)
  044324.

\bibitem{rin80}
P.~Ring, P.~Schuck, The nuclear many-body problem, Springer, Berlin, 1980.

\bibitem{row70}
D.~J. Rowe, Nuclear collective motion, Methuen, London, 1970.

\bibitem{van81}
N.~Van Giai, H.~Sagawa, Nucl. \ Phys. \ A 371 (1981) 1.

\bibitem{roc12}
X.~Roca-Maza, G.~Pozzi, M.~Brenna, K.~Mizuyama, G.~Col\`o, Phys. \ Rev. \ C 85
  (2012) 024601.

\bibitem{hil07}
S.~Hilaire, M.~Girod, Eur. \ Phys. \ J. \ A 33 (2007) 237.

\bibitem{del10}
J.-P. Delaroche, M.~Girod, J.~Libert, H.~Goutte, S.~Hilaire, S.~P\'eru,
  N.~Pillet, G.~F. Bertsch, Phys. \ Rev. \ C 81 (2010) 014303.

\bibitem{co12a}
G.~Co', V.~De~Donno, P.~Finelli, M.~Grasso, M.~Anguiano, A.~M. Lallena,
  C.~Giusti, A.~Meucci, F.~D. Pacati, Phys. \ Rev. \ C 85 (2012) 024322.

\bibitem{ahr75}
J.~Ahrens, et~al., Nucl. \ Phys. \ A 251 (1975) 479.

\bibitem{spe91}
J.~Speth, J.~Wambach, Theory of giant resonances. in Electric and magnetic
  giant resonances in nuclei, {\rm J. Speth ed.}, World Scientific, Singapore,
  1991.

\bibitem{bor98}
P.~F. Bortignon, A.~Bracco, R.~A. Broglia, Giant resonances. Nuclear structure
  at finite temperature, Harwood Academic Press, New York, 1998.

\bibitem{har01}
M.~N. Harakeh, A.~van~der Woude, Giant resonances, Claredon press, Oxford,
  2001.

\bibitem{deh82}
R.~de~Haro, S.~Krewald, J.~Speth, Nucl. \ Phys. \ A 388 (1982) 265.

\bibitem{co85}
G.~Co', S.~Krewald, Nucl. \ Phys. \ A 433 (1985) 392.

\bibitem{tra87}
M.~Traini, G.~Orlandini, R.~Leonardi, Rivista \ Nuovo \ Cimento 10 (1987) 1.

\bibitem{kam04}
S.~Kamerdzhiev, J.~Speth, G.~Tertychny, Phys. \ Rep. 393 (2004) 1.

\bibitem{gam11}
D.~Gambacurta, M.~Grasso, F.~Catara, Phys. \ Rev. \ C 84 (2011) 034301.

\bibitem{lan11}
E.~G. Lanza, A.~Vitturi, M.~V. Andr\'es, F.~Catara, D.~Gambacurta, Phys. \ Rev.
  \ C 84 (2011) 064602.

\bibitem{lan09}
E.~G. Lanza, F.~Catara, D.~Gambacurta, M.~V. Andr\'es, P.~Chomaz, Phys. \ Rev.
  \ C 79 (2009) 054615.

\bibitem{rei12}
P.~G. Reinhard, W.~Nazarewicz, arxiv:1211.1649 [nucl-th] (2012).

\bibitem{ter07}
G.~Tertychny, V.~Tselyaev, S.~Kamerdzhiev, F.~Gr{\"u}mmer, S.~Krewald,
  J.~Speth, A.~Avdeenkov, E.~Litvinova, Phys. \ Lett. \ B 647 (2007) 104.

\bibitem{avd11}
A.~Avdeenkov, S.~Goriely, S.~Kamerdzhiev, S.~Krewald, Phys. \ Rev. \ C 83
  (2011) 064316.

\bibitem{gam12}
D.~Gambacurta, M.~Grasso, V.~De~Donno, G.~Co', F.~Catara, Phys.\ Rev. \ C 86
  (2012) 021304(R).

\end{thebibliography}
\end{document}